\def\dbcol{double column single spacing}
\ifdefined\dbcol
\ifdefined\ACM
	\documentclass{sig-alternate-10pt}
	\makeatletter
	\def\@copyrightspace{\relax}
	\makeatother
\else
	\documentclass[conference,10pt]{IEEEtran}
	\IEEEoverridecommandlockouts
	\makeatletter
	\def\ps@headings{%
	\def\@oddhead{\mbox{}\scriptsize\rightmark \hfil \thepage}%
	\def\@evenhead{\scriptsize\thepage \hfil \leftmark\mbox{}}%
	\def\@oddfoot{}%
	\def\@evenfoot{}}
	\makeatother
\fi

\usepackage[left=.6in,right=.59in,top=.6in,bottom=.5in]{geometry}

\else
	\documentclass[conference,12pt,onecolumn]{IEEEtran}
\fi

	\usepackage[cmex10]{amsmath}
	\usepackage{amsthm}
    \ifdefined\COMSOC
        \usepackage[nocompress]{cite}
    \else
        \usepackage{cite}
    \fi
	\usepackage[pdftex]{graphicx}

\ifdefined\META     \usepackage{stmaryrd}    \fi

\newcommand{\begproof}{\ifdefined\dbcol\begin{IEEEproof}\else\begin{proof}\fi}
\newcommand{\Endproof}{\ifdefined\dbcol\end{IEEEproof}\else\end{proof}\fi}
\newcommand{\metacom}[1]{\ifdefined\META\bluepure{$\blacktriangleright$}#1\bluepure{$\rrbracket$}\fi}
\newcommand{\metafoot}[1]{\ifdefined\META\footnote{\bluepure{$\blacktriangleright$}#1}\fi}

\ifdefined\COMSOC
  \usepackage[font=footnotesize,labelfont=sf,textfont=sf]{subfig}
\else
  \usepackage[font=footnotesize]{subfig}
\fi

\usepackage{amssymb}
\usepackage{bm}
\usepackage{bbm}
\usepackage{verbatim}
\usepackage{color}
\usepackage[normalem]{ulem}
\usepackage[ruled,lined,linesnumbered]{algorithm2e}

\newtheorem{thm}{Theorem}
\newtheorem{lem}{Lemma}
\newtheorem{prop}{Proposition}

\newtheorem{defn}{Definition}

\newcommand{\fref}[1]{Fig.~\ref{#1}}

\newcommand{\sref}[1]{Section~\ref{#1}}
\newcommand{\thmref}[1]{Theorem~\ref{#1}}
\newcommand{\lref}[1]{Lemma~\ref{#1}}
\newcommand{\pref}[1]{Proposition~\ref{#1}}
\newcommand{\cref}[1]{Corollary~\ref{#1}}

\newcommand{\vect}[1]{\boldsymbol{\mathbf{#1}}}

\newcommand{\blue}[1]{{\color{blue}\dotuline{#1}}}
\newcommand{\bluepure}[1]{{\color{blue}{#1}}}

\newcommand{\nn}{\nonumber}

\newcommand{\opd}{\operatorname{d}\!}

\newcommand{\pd}[2]{\frac{\partial #1}{\partial #2}}

\newcommand{\inv}[1]{\frac{1}{#1}}
\DeclareMathOperator*{\argmax}{arg\,max}

\usepackage{enumitem}

\usepackage{datetime}

\usepackage{fancyhdr}
\fancypagestyle{empty}{
  \fancyhf{}
  \fancyhead[C]{\footnotesize{T. Luo, S. S. Kanhere, S. K. Das and H-P. Tan, ``Optimal prizes for all-pay contests in heterogeneous crowdsourcing'', {\it Proc. IEEE MASS}, 2014, pp. 136-144.}}
}

\title{Optimal Prizes for All-Pay Contests in \\Heterogeneous Crowdsourcing}
\author{
  \IEEEauthorblockN{Tie Luo\IEEEauthorrefmark{1}, Salil S. Kanhere\IEEEauthorrefmark{2}, Sajal K. Das\IEEEauthorrefmark{3}, Hwee-Pink Tan\IEEEauthorrefmark{1}
  \IEEEauthorblockA{
	\IEEEauthorrefmark{1}Institute for Infocomm Research, A*STAR, Singapore\\
	\IEEEauthorrefmark{2}School of Computer Science and Engineering, The University of New South Wales, Australia\\
	\IEEEauthorrefmark{3}Department of Computer Science, Missouri University of Science and Technology, USA\\
    E-mail: luot@i2r.a-star.edu.sg, salilk@unsw.edu.au, sdas@mst.edu, hptan@i2r.a-star.edu.sg}
} }

\begin{document}
\maketitle
\thispagestyle{empty}

\begin{abstract}
Incentive is key to the success of crowdsourcing which heavily depends on the level of user participation. This paper designs an incentive mechanism to motivate a heterogeneous crowd of users to actively participate in crowdsourcing campaigns. We cast the problem in a new, asymmetric {\em all-pay contest} model with incomplete information, where an arbitrary $n$ of users exert irrevocable effort to compete for a {\em prize tuple}. The prize tuple is an array of prize functions as opposed to a single constant prize typically used by conventional contests. We design an optimal contest that (a) induces the {\em maximum profit}---total user effort minus the prize payout---for the crowdsourcer, and (b) ensures users to {\em strictly} have incentive to participate. In stark contrast to intuition and prior related work, our mechanism induces an equilibrium in which heterogeneous users behave {\em independently} of one another as if they were in a homogeneous setting. This newly discovered property, which we coin as {\em strategy autonomy} (SA), is of practical significance: it (a) reduces computational and storage complexity by $n$-fold for {\em each} user, (b) increases the crowdsourcer's revenue by counteracting an {\em effort reservation} effect existing in asymmetric contests, and (c) neutralizes the (almost universal) law of diminishing marginal returns (DMR). Through an extensive numerical case study, we demonstrate and scrutinize the superior profitability of our mechanism, as well as draw insights into the SA property.
\end{abstract}


\section{Introduction}\label{sec:intro}

Crowdsourcing offers a cost-effective approach to distributed problem solving and data collection by  soliciting contributions (solutions, ideas, data, etc.) from a large group of people. Compared to conventional means of hiring employees, crowdsourcing can be potentially more cost-efficient. It has thus catalyzed new computing and sensing paradigms such as 
participatory sensing\cite{ps10mobisys}.

Key to the viability of crowdsourcing is providing {\em incentives} to attain a sufficient level of user participation. While incentives can be classified into intrinsic motivation (e.g., self-fulfillment, enjoyment, and esteem) and extrinsic incentives (e.g., peer pressure, financial rewards) \cite{kreps97}, we focus on monetary incentives which fall in the second category and have wider applications in practice. Specifically, in this paper, we employ the theory of mechanism design, in particular {\em auctions}\cite{krishna09}, to design an incentive mechanism for crowdsourcing.

We choose auction theory to be the tool primarily because, to reward users for their contributions, a unilaterally stipulated pricing scheme---say by some authority---for such informational products as ideas, data, solutions, etc., is hard to satisfy both the crowdsourcer and users. On the other hand, auctions offer a perfect mechanism for a {\em principal} (crowdsourcer) and {\em agents} (users) to implicitly negotiate and mutually agree on a deal (e.g., solving a problem, contributing data). Indeed, auctions as a bilateral pricing scheme have been widely adopted by a sizable body of literature \cite{hoh10percom,yang12mobicom,kout13infocom,luo14infocom}.

In this paper, we design an incentive mechanism by modeling crowdsourcing as an {\em all-pay contest}\cite{siegel09}. All-pay contests are {\em isomorphic} to all-pay auctions\cite{krishna09}---given an equilibrium in one model, one can construct one and only one equilibrium in the other model\footnote{Drawing an analogy may shed light on the isomorphism: contestants exerting effort in a contest is like bidders tendering bids in an auction, and the highest-effort contestant winning the prize is like the highest bidder securing the auctioned good.}---yet contests are semantically closer to our mechanism which uses prizes as auctioned goods. Auctions have many flavors in which the most common and well-known ones, such as English (or second-price) and Dutch (or first-price) auctions, belong to the intuitive ``winner-pay'' genre---only the winner (i.e., the highest bidder) pays for a bid.
On the contrary, ``all-pay'' auctions require {\em every} bidder to pay for his bid regardless of who wins the auction. This seemingly peculiar form, however, precisely describes most crowdsourcing platforms (e.g., Amazon mechanical turk (AMT), task.cn, TaskRabbit.com) in which there is only one winner but all participants will have to exert their {\em irrevocable} effort {\em before} the winner is announced. This is why we model crowdsourcing as an all-pay auction or contest.

The vast majority of prior work using auctions (either all-pay or winner-pay) \cite{SODA12,dv09csallpay,AS09ICIS,yang12mobicom,kout13infocom,luo14infocom} is based on {\em symmetric} auctions where bidders are {\em ex post} or {\em ex ante} identical, i.e., either they have exactly the same type\footnote{In Bayesian games and mechanism design, {\em type} is a term that refers to an agent's private information or signal (e.g., ability, skill level, valuation of the auctioned good), which essentially characterizes an agent.} (ex post), or their types follow the same, single probabilistic distribution (ex ante). As a result, they all model an environment of {\em homogeneous} agents, which is amicable to analysis. However, in many real crowdsourcing scenarios (e.g., AMT, task.cn, TaskRabbit.com, TopCoder.com), users come from dissimilar backgrounds and possess different skill levels, constituting a {\em heterogeneous} environment. Therefore, a more realistic model would be {\em asymmetric} auctions. Unfortunately, asymmetric auctions are much less studied and understood because their associated analysis is much more challenging, due to the fact that the most celebrated {\em revenue equivalence theorem}\footnote{This theorem states that, under a set of standard assumptions including symmetric agents, any auction generates the same revenue for the principal.} \cite{myerson81,RS81AER} breaks under asymmetry. Thus, as a compromise for analytical tractability, most research on asymmetric auctions is limited to two-player cases\cite{maskin00asym,amann96asym,nora14asym} or {\em complete-information} settings\cite{siegel09,xiao14asym} in which all types are commonly known. A significant progress was made by \cite{fibich03asym} toward understanding asymmetric {\em first-price} auctions with multiple players and incomplete information, but only an {\em approximate} solution was obtained and it only applies to {\em weakly} asymmetric agents. Other studies in this context resort to numerical methods\cite{fibich11simu}.

Unlike prior work, our model accommodates {\em asymmetric}---regardless of weak or strong---all-pay contests with an arbitrary $n$ of players and {\em incomplete} information (i.e., agent types are uncertain). This is a much closer characterization of most real crowdsourcing scenarios. Furthermore, we obtain precise, analytical solutions rather than approximate or numerical ones. Thus, this work presents the first attempt to offer a rigorous understanding toward more realistic crowdsourcing campaigns, with results of a much wider applicability. This constitutes the {\bf first contribution} of this paper.

The {\bf second contribution} is that we explore another degree of freedom in contest design, by furnishing the contest with a {\em prize tuple} that consists of an array of $n$ prize {\em functions} to cater for the heterogeneous agents. This is distinct from all the standard contests or auctions where a single {\em fixed} prize or good is used. The rationale for this exploration is to understand {\em whether} and {\em how} such a prize tuple can elicit higher  {\em profit}---revenue (agent effort) minus cost (prize payout)---for the principal. In this paper, we derive the optimal prize tuple under asymmetry, and demonstrate that it can induce significantly higher profit than both symmetric and asymmetric fixed-prize contests.

The {\bf third contribution} is that this paper is the first to incorporate the {\em principal's valuation of the prize} into profit formulation, thereby enabling our model and results to apply to different crowdsourcers. It also means that, even if a contest adopts a fixed prize, the {\em cost} (valuation of the prize) can still be variable. Therefore, regardless of fixed or variable prizes, this work provides an example of how to cover different crowdsourcers or the same crowdsourcer's varying valuations.

Our {\bf fourth contribution}, which we stress, is the discovery and investigation of a new and counter-intuitive property pertaining to asymmetric auctions and contests, called {\em strategy autonomy} (SA). It captures the phenomenon that agents in an asymmetric equilibrium behave {\em independently} of one another as if they were in a {\em symmetric} one. This is in stark contrast to all prior work on asymmetric auctions, and has three practical significances: it (a) reduces computational and storage complexity from $O(n)$ to $O(1)$ for {\em each} agent, (b) increases principal's revenue by counteracting an {\em effort reservation} effect engendered by asymmetric belief, and (c) dramatically enhances the system scalability by neutralizing the (almost universal) law of diminishing marginal returns (DMR). Moreover, in addition to SA, our mechanism also strictly satisfies {\em individual rationality} (IR), which means that all agents strictly have incentive to participate in our mechanism.

The rest of this paper proceeds as follows. \sref{sec:relwk} reviews the literature. \sref{sec:model} presents our model and \sref{sec:analysis} provides the analysis. A detailed numerical case study is then given in \sref{sec:simu} which demonstrates key results, offers intuitions, as well as draw some insights. \sref{sec:conc} concludes the paper.

\section{Related Work}\label{sec:relwk}

\subsection{Incentive mechanisms and Symmetric auctions}
While auctions are widely recognized as an effective incentive mechanism for various activities such as online sales, government procurement, crowdsourcing and participatory sensing, the vast majority of prior work on auction-based incentive presumes symmetric auctions. For example, in \cite{yang12mobicom} and \cite{kout13infocom}, users bid their costs and the crowdsourcer or service provider determines their payments such that the users bid their costs truthfully. Along a different spirit, \cite{luo14infocom} investigates an {\em all-pay auction} based incentive mechanism that is tailored to realistic settings, including uncertain population sizes, unknown (yet symmetric) user types, and risk-averse (subsuming risk-neutral) users.

Other studies on optimal design of auctions or contests consider different approaches and objectives. For example, \cite{MS01AER,AS09ICIS} investigate whether a single or multiple prizes is optimal in terms of maximizing the highest $k$ bids\cite{AS09ICIS} or total bids \cite{MS01AER}. Under a similar setting, \cite{SODA12} shows that the highest bid is at least half of the total bids, and \cite{cohen08} finds that there is no advantage to have multiple prizes under certain conditions. Similarly in a symmetric model, \cite{kaplan02} introduces variable rewards and examines a paradoxical behavior where a reduction in reward or an increase in cost may increase the total or the highest bid in expectation.

Indeed, almost all auction theory, including the above, concerns symmetric auctions, as concurred by \cite{fibich12dyn}. On the contrary, this paper tackles the challenge of asymmetry in order to accommodate heterogeneous and more realistic crowdsourcing environments.

\subsection{Asymmetric auctions}
This domain is relatively much less understood due to its analytical complexity. Most work in this domain is devoted to two-player cases or complete-information settings for the sake of tractability. Amann and Leininger's seminal work\cite{amann96asym} offers an analysis of the equilibrium strategies for a two-player asymmetric case. It was then extended by Maskin and Riley\cite{maskin00high,maskin03uniq} who proved the monotonicity and uniqueness of the equilibrium. Unfortunately, till now, there still lacks a closed-form solution to general $n$-player cases with incomplete information.
Fibich and Gavious\cite{fibich03asym} proposed a perturbation approach to obtain an analytical, but approximate, solution to equilibria on the premise of weak asymmetry. Improvement in terms of some other mathematical properties was made later using a dynamic-system approach\cite{fibich12dyn}. Another work \cite{PR10hetero} on asymmetric contests focuses on risk aversion and gives some exploratory yet inconclusive results.

Other studies assume complete information. Siegel\cite{siegel09} was probably the first who coined the term ``all-pay contests'', where he analyzed closed-form player payoffs in equilibrium with complete information. Under a similar model, Xiao \cite{xiao14asym} studied the problem of allocating more than one prizes to a number of winners and proposed an algorithm to construct the equilibrium. Franke et al.\cite{franke13asym} aimed to maximize the revenue (which is part of the profit we study in this paper) through discriminating players by associating differentiated weights to players, assuming complete information is available.

Moreover, all the above studies presume fixed prizes.

In this paper, we allow for more generality by not assuming the availability of complete information and by accommodating multiple asymmetric players. Secondly, we empower the crowdsourcer to provision a tuple of individually different prize functions, in order to induce the highest possible profit. All these are clearly different from prior art.

\section{The Model}\label{sec:model}

The overall problem setting is that a principal aims to crowdsource from $n$ heterogeneous agents for the maximal profit---total effort (revenue) minus prize (cost). In this paper, ``effort'' is a general term that can be interpreted according to different contexts; for example, it can refer to the quantity of tasks completed, the quality of solutions submitted\cite{luo14secon}, or a quality-modulated quantity of sensing data\cite{luo13qcs}.

We design this crowdsourcing campaign under an {\em all-pay contest} framework, in which all participating agents exert effort (which is irrevocable and will be sunk, i.e., ``all-pay'', regardless of the outcome) in order to win some prize (i.e., as if in a ``contest'').  An agent $i=1,2,...,n$ is characterized by his {\em type} $v_i$ which is private information (e.g., skillfulness or competency): agent $i$ knows his own type $v_i$ but does not know any other $v_j, \forall j\ne i$. On the other hand, it is common knowledge that all the agent types $v_i|_{i=1}^n$ are independently drawn from $[\uline v, \bar v]$ according to $F_i(v)|_{i=1}^n$, respectively, where $F_i(v)$ are the c.d.f. of $v_i$ and $[\uline v, \bar v]$ is a continuous and nonnegative support.  This setting corresponds to an {\em incomplete-information} setting and, essentially, constitutes an {\em asymmetric} Bayesian game where the common prior consists of $n$ generally different distributions. Without loss of much generality, we assume that each $F_i(v)$ is differentiable and the corresponding p.d.f. $f_i(v)$ is continuous and positive over $(\uline v, \bar v)$.

To provide incentive, the principal will reward the agent who exerts the highest effort, i.e., the ``winner'', a prize. In this paper, we allow the prize to depend on the winner's effort rather than fixing the prize ex ante. Furthermore, in view of the asymmetric agents, we provision the prize as a {\em prize tuple} $\vect Z:=\langle Z_1(b_1),Z_2(b_2),...,Z_n(b_n)\rangle$, where $Z_i(b_i)$ is a prize function in agent $i$'s effort $b_i$ and takes effect when $i$ is the winner.
This prize tuple is known to all the agents before the contest.

The value of a prize to an agent is characterized by a value function $V(v,Z)$: an agent of type $v$ values a prize $Z$ (the face value) to be of real worth $V(v,Z)$ to him. That is, agent $i$ values the prize he competes for to be of real value $V(v_i, Z_i(b_i))$. An example of the value function is $V(v,Z)=v Z$, meaning that an agent with higher ability can gain more benefit out of a prize; in fact, $V(\cdot)=v Z$ is a generalization of standard all-pay auctions where the fixed auctioned good is normalized and $V(\cdot)=v$.
With out loss of much generality, we assume that $V(v,Z)$ is differentiable with respect to $v$. 

Exerting effort incurs cost; an agent $i$ who exerts effort $b_i$ has to pay for his cost as per a payment function $p(b_i, v_i)$. We do not assume a specific form for $p(\cdot)$ or $V(\cdot)$, thereby covering a broad class of auctions and contests.\footnote{For example, letting $p(b,v)=b$ and $V(v,Z(b))=v$ yields standard all-pay auctions;
letting $p(b,v)=0$ and $V(v,Z(b))=v-b$ yields standard first-price auctions.
} Now we can formulate the expected utility of agent $i$, as
\begin{align}\label{eq:util-def-original}
u_i := q_i V(v_i,Z_i(b_i)) - p(b_i,v_i)
\end{align}
where $q_i$ is the probability that agent $i$ wins the contest.

The principal aims to maximize his expected profit $\pi$, defined as the total crowdsourced effort minus his prize payout, or formally,
\begin{align}\label{eq:profit-def}
\pi := \mathbbm E\big[\sum_i b_i - V(\lambda, Z_w(b_w)) \big] 
\end{align}
where $w\in[1...n]$ is the winner's index which is a random variable, and $\lambda>0$ is the principal's type (valuation of prize) which is common knowledge. In this profit formulation, the revenue is defined in terms of {\em total} effort, which covers data-gathering kinds of applications; it can also be defined in terms of the {\em highest} effort, which covers solution-eliciting kinds of applications, but the corresponding analysis completely parallels this paper.

Throughout, we follow the notation convention of $g'_x:=\pd{g}{x}$, $g''_{xy}:=\frac{\partial^2 g}{\partial x \partial y}$, and $g'''_{x^2 y}:=\frac{\partial^3 g}{\partial x^2 \partial y}$, for any differentiable function $g(x,y)$.  We assume that the payment function $p(b,v)$ is twice continuously differentiable, $p(0,v)=0$, $p'_b(b,v)>0$ which means higher effort, higher payment (cost); $p'_v(b,v)\le 0$ which means higher type (ability), lower payment; $p''_{b b}(b,v)>0$ which means striving from higher effort levels is more costly than from lower effort levels, or conversely, the marginal output by adding effort is decreasing; finally, $p'''_{b^2 v}(b,v)\le 0$ which means lower types are more vulnerable to the decreasing marginal output. We note that these assumptions are not restrictive.

In practice, our mechanism modeled in this section can be realized as follows. The agent type distributions $F_i$ can be constructed by the crowdsourcer from user contribution history, or from social acquaintance in the case of a small local community of users, and then published on a website or via a mobile app. The prize tuple can be either published or downloaded, where each user has software (e.g., a mobile app) to act as his agent. The user effort can be measured (in terms of time, data volume, sampling rate, etc.) (a) by each agent itself, or (b) by the principal (e.g., in a cloud) and continually fed back to each corresponding agent, so that the agent can decide how much effort to contribute.

\section{Analysis}\label{sec:analysis}

We first analyze the asymmetric equilibrium strategy for each agent (\lref{lem:strategy}), which is a function of any given prize tuple. Then we determine the optimal prize tuple that induces the maximum profit for the principal (\thmref{thm:opt-prize}). Following that is an exposition of three important properties (\sref{sec:property}).

\subsection{Equilibrium strategy}

\begin{defn} [Bayesian Nash equilibrium]
A pure-strategy Bayesian Nash equilibrium is a strategy profile $\vect b^* := (b_1^*, b_2^*,...,b_n^*)$ that satisfies 
\[ u_i(b_i^*,b_{-i}^*) \ge u_i(b_i,b_{-i}^*), \forall b_i, \forall i. \]
\end{defn}
In words, each agent in a Bayesian Nash equilibrium plays a strategy that maximizes his expected payoff given his belief about other agents' types and that other agents play their respective equilibrium strategies.

\begin{lem}[Existence, monotonicity, uniqueness, and common support]\label{lem:exists}
Our asymmetric all-pay contest with incomplete information admits a unique, asymmetric, pure-strategy Bayesian Nash equilibrium that is strictly monotonic, i.e., the equilibrium bids are strictly increasing in type. Furthermore, given that the agent types have a common and nonnegative support, i.e., $[\uline v, \bar v]$, the equilibrium bids also have a common support $[0, \bar b]$, where $\bar b$ is unknown.
\end{lem}
We defer the proof to \cite{mass14app} due to space constraint.

{\bf Notation convention:}
Henceforth, we will exclusively deal with the equilibrium state. Hence for brevity, we slightly deviate from the general national convention
by dropping the superscript $^*$ on equilibrium variables. For example, we write $b_i$ instead of $b_i^*$ and $v_i(\cdot)$ instead of $v_i^*(\cdot)$.

\lref{lem:exists} tells that an agent's equilibrium strategy $b_i$ is a strictly monotone (increasing) function of $v_i$, which we denote by $\beta_i(\cdot)$, i.e., $b_i=\beta_i(v_i)$. Thus, its inverse function exists and is also increasing, which we denote by $v_i(\cdot):=\beta_i^{-1}(\cdot)$. Moreover, because of the strict monotonicity, the event $b_i=b_j$ is of zero probability and tie-breaking is trivial. Thus,
\[ \Pr(b_i>b_j)=\Pr(\beta_j^{-1}(b_i) > v_j) =F_j(v_j(b_i) ). \]
Furthermore, because agent $i$'s winning probability $q_i=\prod_{j\ne i} \Pr(b_i>b_j)$, \eqref{eq:util-def-original} can be rewritten as
\begin{align}\label{eq:util-def}
u_i = V(v_i,Z_i(b_i)) \prod_{j\ne i}F_j(v_j(b_i)) - p(b_i,v_i).
\end{align}

\begin{lem}\label{lem:strategy}
Given a prize function $Z_i(\cdot)$, an agent $i$'s equilibrium strategy $b_i(v_i)$ is determined by \metafoot{This result should tally with \cite{amann96asym} if setting $p()=b$. The assumption on convexity of $p$ is only relevant to Theorem 1 (profit maximization). - but I didn't have time to verify this.}
\begin{multline}\label{eq:strategy}
V(v_i,Z_i(b_i)) \prod_{j\ne i}F_j(v_j(b_i)) - p(b_i,v_i) \\
= \int_{\uline v}^{v_i} [ V'_{v_i} (\tilde v_i,Z_i(b_i)) \prod_{j\ne i}F_j(v_j(b_i)) - p'_{v_i} (b_i,\tilde v_i) ] \opd \tilde v_i.
\end{multline}
\end{lem}\metafoot{\lref{lem:strategy} holds regardless of our assumptions on $p()$.}
\begproof
As the equilibrium bid $b_i$ is also the solution to the optimization problem $\max_{b_i} \{u_i\}$ \eqref{eq:util-def}, we invoke the envelope theorem\cite{envelope02} on \eqref{eq:util-def} with respect to $v_i$ and obtain
\begin{align}
\pd{u_i}{v_i} &= V'_{v_i}( v_i,Z_i(b_i)) \prod_{j\ne i} F_j(v_j(b_i)) - p'_{v_i} (b_i,v_i) \nn\\
\Rightarrow u_i(v_i) &= u_i(\uline v) \nn\\
+\int_{\uline v}^{v_i} &\Big[V'_{v_i} (\tilde v_i,Z_i(b_i)) \prod_{j\ne i}F_j(v_j(b_i)) 
- p'_{v_i} (b_i,\tilde v_i) \Big] \opd \tilde v_i. \label{eq:ut-expand}
\end{align}
\metafoot{It often happens that $\uline v$ is not in the support of the integrand (e.g. containing $1/v$). In that case the above could be changed to $u_i(v_i) = u_i(\bar v) - \int_v^{\bar v}$ but the problem then is $u_i(\bar v)$: although [maskin00asym] proves a common $\bar b$ in asymmetric environment, the value of $\bar b$ is unknown. So the compromise we have to make is to stick to $u(\uline v)$ but, when it comes to numerical result, choose cost $p(\cdot)$ such that the integrand is valid at $\uline v$.}
Since an agent with the lowest possible type never wins the auction, he will bid zero (i.e., exert no effort) in an all-pay auction (rather than bidding $b_i=\uline v$ as in first or second-price auctions). As a result, he reaps zero utility, i.e., $u_i(\uline v)=0$. Then, equating the r.h.s of \eqref{eq:ut-expand} to that of \eqref{eq:util-def} yields the result.
\Endproof
Recall in \eqref{eq:strategy} that the derivatives in the integrand denote partial derivatives, and hence no further reduction is allowed.

{\bf Remark:} 
Asymmetric auctions, regardless of winner-pay or all-pay, do not have closed-form expressions for equilibrium strategies in general (while an approximate solution to the first-price flavor can be found in \cite{fibich03asym}). However, we can solve for the optimal prize tuple using \lref{lem:strategy} without obtaining the closed form of equilibrium strategies, as shown next.

\subsection{Optimal prize tuple}\label{sec:opt-prize}

Solving for the optimal, i.e., profit-maximizing, prize tuple $\vect Z$ requires an explicit form of the value function $V(\cdot)$, for which we consider $V(v,Z) := h(v) Z$ where $h(\cdot)$ satisfies $h(0)=0$ and $h'(v)>0$.\metafoot{If allow $h'(v)=0$, first need to add $h(v)>0$ for $v>0$ here; but most importantly, strict IR will not hold (although IR still holds), because \pref{prop:IR} will have possibility of $u_i=0$ for any $v_i$.} This form further generalizes  the form $V=v Z$ which, as mentioned in \sref{sec:model}, is already a generalization of the standard all-pay auctions.

First note that the utility maximization problem of each agent, $\max\limits_{b_i} \{u_i\}$ \eqref{eq:util-def}, can be reformulated as $\argmax_{b_i} \{u_i\}$ without change in principle, i.e.,
\[ \argmax_{b_i} h(v_i) Z_i(b_i) \prod_{j\ne i}F_j(v_j(b_i)) - p(b_i, v_i). \]
It is equivalent to
\begin{align}\label{eq:ui-def-eqv}
\argmax_{b_i}Z_i(b_i) \prod_{j\ne i}F_j(v_j(b_i)) - \hat p(b_i, v_i)
\end{align}
where $\hat p(b, v)=p(b, v)/h(v)$ for $v>0$.\metafoot{This reformulation retains the same solution but will give us convenience in subsequent mathematical manipulations (as the value function $V(v,Z(b))$ becomes $Z(b)$ and hence type-independent).}

\begin{thm}\label{thm:opt-prize}
The optimal prize tuple that maximizes the principal's profit is given by $\vect Z=\langle Z_1(b_1),Z_2(b_2),...,Z_n(b_n)\rangle$ where
\begin{align}\label{eq:opt-prize}
Z_i(b_i) = \frac{ \hat p(b_i, v_i(b_i)) - \int_0^{b_i} \hat p'_{v_i}(\tilde b_i, v_i(\tilde b_i)) \opd v_i(\tilde b_i) }
{ \prod_{j\ne i}F_j(v_j(b_i)) }
\end{align}
in which the optimal effort $b_i(v_i)$ is implicitly given by
\begin{align}\label{eq:opt-effort}
\hat p'_{b_i}(b_i, v_i)  = \inv{h(\lambda)} + \hat p''_{b_i,v_i} (b_i, v_i) \frac{1 - F_i}{f_i}
\end{align}
where $v_i(b_i)$ is the inverse function of $b_i(v_i)$. The maximum profit achieved is given by
\begin{align}\label{eq:profit-final}
\!\pi\!=\!\sum_i \! \int_{\uline v}^{\bar v}\!\!\Big[ b_i(v_i) 
- h(\lambda) \hat p(b_i, v_i) + h(\lambda) \hat p'_{v_i} (b_i(v_i), v_i) \frac{1 - F_i}{f_i} \Big]\! \opd F_i.
\end{align}
\end{thm}
\begproof
First, we expand the principal's expected profit \eqref{eq:profit-def} by calculating the expected cost, i.e., prize. Noticing that an agent $i$'s winning probability is $q_i=\prod_{j\ne i}F_j(v_j(b_i))$, we use the law of total expectation to have
\[  \mathbbm E_w [Z_w(b_w)] = \sum_i \int_{\uline v}^{\bar v} Z_i(b_i(v_i)) \prod_{j\ne i} F_j(v_j(b_i)) \opd F_i (v_i). \] 
Then, by expanding the revenue portion,
\begin{align}\label{eq:profit-expand}
\pi &= \Big[ \sum_i \int_{\uline v}^{\bar v} b_i(v_i) \opd F_i(v_i) \Big]
- h(\lambda) \mathbbm E_w [Z_w(b_w)] \nn\\
&= \sum_i \int_{\uline v}^{\bar v} \Big[ b_i(v_i) 
- h(\lambda) Z_i(b_i(v_i)) \prod_{j\ne i} F_j(v_j(b_i)) \Big] \opd F_i(v_i).
\end{align}

With \lref{lem:strategy}, we substitute $Z_i(b_i)$ for $V(v_i,Z_i(b_i))$ in \eqref{eq:strategy} 
and $\hat p(b_i, v_i)$ for $p(b_i, v_i)$, and obtain
\begin{align}\label{eq:strategy-expand}
\hspace{-2.5mm}Z_i(b_i) \prod_{j\ne i}F_j(v_j(b_i)) - \hat p(b_i, v_i) 
= - \int_{\uline v}^{v_i} \hat p'_{v_i}(b_i(\tilde v_i), \tilde v_i) \opd \tilde v_i,
\end{align}
where note that ${Z_i}'_{v_i}(b_i)=0$ due to the envelope theorem. 
Substituting \eqref{eq:strategy-expand} into \eqref{eq:profit-expand} yields
\begin{align*}
\!\pi\! =\! \sum_i\! \int_{\uline v}^{\bar v}\!\! \Big[ b_i(v_i) 
- h(\lambda) \hat p(b_i, v_i) + h(\lambda)\! \int_{\uline v}^{v_i}\! \hat p'_{v_i}(b_i(\tilde v_i), \tilde v_i) \opd \tilde v_i \Big]\! \opd F_i.
\end{align*}
By integrating the last term by parts,
\begin{align*}
&\quad \int_{\uline v}^{\bar v} \int_{\uline v}^{v_i} \hat p'_{v_i}(b_i(\tilde v_i), \tilde v_i) \opd \tilde v_i \opd F_i \\
&= \int_{\uline v}^{\bar v} \hat p'_v (b_i(v_i), v_i) \opd v_i
- \int_{\uline v}^{\bar v} F_i(v_i) \hat p'_{v_i} (b_i(v_i), v_i) \opd v_i \\
&= \int_{\uline v}^{\bar v} \hat p'_{v_i} (b_i(v_i), v_i) \frac{1 - F_i}{f_i} \opd F_i.
\end{align*}
Substituting this back proves the principal's profit \eqref{eq:profit-final}.

The principal originally faces the problem of $\max_{\vect Z} \{\pi\}$, yet it is equivalent to $\max_{\vect b} \{\pi\}$ because, essentially, the prize tuple is used to induce the optimal effort vector $\vect b$.  Furthermore, in \eqref{eq:profit-final} we have decoupled each agent $i$ from other agents $j\ne i$. Therefore, maximizing $\pi$ can be achieved by maximizing each individual integrand $I_i$ over $b_i$, where
\[ I_i := b_i(v_i) - h(\lambda) \hat p(b_i, v_i) 
+ h(\lambda) \hat p'_{v_i} (b_i, v_i) \frac{1 - F_i}{f_i}. \]
Simply applying the first order condition to $I_i$ with respect to $b_i$ proves the optimal effort $b_i$ \eqref{eq:opt-effort} for each agent $i$. 

To verify that the above effort $b_i$ is the unique maximizer, we examine the second derivative
\[ {I_i}''_{b_i^2} = - h(\lambda) \hat p''_{b_i^2}(b_i, v_i) 
+ h(\lambda) \hat p'''_{b_i^2 v_i} (b_i, v_i) \frac{1 - F_i}{f_i}. \]
Since $\hat p=p/h(v)$, and $v>0$ is treated as a fixed value due to the envelope theorem, our assumptions on $p(\cdot)$ also hold for $\hat p(\cdot)$, i.e., $\hat p''_{b_i^2}>0$ and $\hat p'''_{b_i^2 v_i}\le 0$. Since $h(\lambda)>0$ for $\lambda>0$, therefore $I_i''<0$. Thus $I_i$ is strictly concave, which validates the existence and uniqueness of $b_i$ given by \eqref{eq:opt-effort}.

The optimal prize function \eqref{eq:opt-prize} is then obtained, by rearranging \eqref{eq:strategy-expand} and changing variables from $v_i$ to $b_i$. Note that the lower limit of integral, 0, is due to $b_i(\uline v)=0$ as the lowest-type agent will bid zero (cf. 
proof of \lref{lem:strategy}).
\Endproof

\subsection{Qualitative properties}\label{sec:property}

This section states three properties pertaining to our mechanism, denoted by OPT since it is equipped with an optimal prize tuple.

\vspace{1mm}
\subsubsection{Strategy Autonomy (SA)}
This is perhaps the most salient property of OPT, particularly in the presence of asymmetry. SA is of practical significance and none of prior work on asymmetric mechanisms possesses this property.
\begin{defn} [Strategy Autonomy]
A mechanism satisfies strategy autonomy if,
given that the common prior $F_i(v_i)|_{i=1}^n$ are individually different, the equilibrium strategy $b_i(v_i | F_{-i}) = b_i(v_i), \forall i$.
\end{defn}
In words, in an asymmetric incomplete-information setting where agents are ex ante heterogeneous, each agent adopts in equilibrium a strategy that is {\em independent} of (the prior about) the other agents. In other words, despite an asymmetric environment (belief), agents behave {\em autonomously} as if they were in a symmetric one.

\begin{prop}
The OPT mechanism satisfies strategy autonomy.
\end{prop}
\begproof
Immediately follows from \thmref{thm:opt-prize}, where the equilibrium strategy $b_i$ \eqref{eq:opt-effort} is independent of any $j\ne i$.
\Endproof

This is a rather counter-intuitive, and somewhat surprising result. This is because \lref{lem:strategy} shows that the equilibrium strategy $b_i$ {\em does} depend on $F_j|_{j\ne i}$, or the strategy is {\em not} autonomous, which also conforms to our intuition as the environment is asymmetric. Indeed, SA is in direct opposition to all prior work on asymmetric auctions, regardless of winner-pay or all-pay, with complete or incomplete information; see, e.g., \cite{amann96asym,fibich03asym,xiao14asym} and a comprehensive survey \cite{konrad09book}.
So the key question is: {\em why} do agents behave autonomously in the OPT mechanism?

The fundamental reason is that the asymmetric belief about agent types is {\em endogenized} (i.e., ``absorbed'') by the optimal prize functions \eqref{eq:opt-prize}, or in other words, any bidder $i$'s reasoning about other bidders' bids is implicitly captured by the function $Z_i(\cdot)$. The rationale of this, i.e., isolating asymmetry from agents, is to counteract an {\em effort reservation} effect arising from asymmetric belief, which is explained below.

SA has three important practical implications:

$\bullet$ {\it Reduces complexity and saves energy:} SA remarkably reduces the computational complexity and storage requirement, from $O(n)$ to $O(1)$, for {\em each} agent. The $O(n)$ can be understood from \eqref{eq:strategy} where each agent's strategy involves reasoning about all the $j\ne i$, which is also the case in, e.g., \cite{fibich03asym,fibich12dyn,franke13asym}.  This advantage enables each agent, which is embodied in practice typically by software that resides on distribute portable devices (e.g., mobile phones), to shed substantial computational and storage burden and, as a result, save considerable energy.

$\bullet$ {\it Counteracts effort reservation:} SA overcomes an effort reservation effect that exists in standard (fixed-prize) asymmetric auctions\cite{cant08asymrev}: when the prize is fixed, any agent only needs to win the other agents by an infinitesimal {\em winning margin}; therefore, using a two-player scenario to illustrate, if the stronger agent believes that the other agent is statistically weaker, he has the incentive to reserve effort in order to reduce his winning margin since a larger margin does not make the winner better off at all. This effect outweighs the strategy adjustment of the weaker agent and results in a reduced total revenue compared to symmetric auctions\cite{cant08asymrev}, which will also be demonstrated in \sref{sec:simu}. However, SA insulates such negative inter-agent influence, allowing agent not to be concerned with other agents and to concentrate on exerting higher effort to increase the winning margin which is now qualified for the (variable) prize.

$\bullet$ {\it Neutralizes the law of DMR}: The prevailing law of diminishing marginal returns (DMR) governs many phenomena in (network) economics. It states that, as the number of new employees increases, the marginal product of an additional employee will at some point be less than the marginal product of the previous employee\cite{microecon01}. Mathematically, DMR leads to a {\em concave} growth of profit or revenue as employees are being added, which is also demonstrated by our evaluation of a standard (fixed-prize) auction in \sref{sec:simu} (\fref{fig:nsym}). However, the independence connoted by SA neutralizes this submodularity-resembling law of DMR, and indeed, we will show in \sref{sec:simu} that the principal's profit increases {\em linearly} as the number of agents increases. This translates to a dramatically enhanced system {\em scalability}.

\subsubsection{Individual Rationality (IR)} 
\begin{defn}[Individual Rationality]
A mechanism satisfies individual rationality if, in equilibrium, all participating agents expect (weakly) higher surplus than from not participating. That is, $u_i(b_i, b_{-i}) \ge u_i(0, b_{-i})$ for all $i$ in equilibrium.
\end{defn}
In other words, a mechanism satisfying IR ensures that any agent has incentive to participate.

\begin{prop}\label{prop:IR}
The OPT mechanism satisfies individual rationality. In particular, an agent reaps {\em strictly} positive utility if he exerts positive effort.
\end{prop}
\begproof
Combining \eqref{eq:strategy-expand} and \eqref{eq:ui-def-eqv} yields
\begin{align*}
\frac{u_i}{h(v_i)} &= - \int_{\uline v}^{v_i} \hat p'_{v_i}(b_i(\tilde v_i), \tilde v_i) \opd \tilde v_i \\ \Rightarrow
u_i &= - h(v_i) \int_{\uline v}^{v_i} 
\frac{p'_{v_i}(b_i, \tilde v_i) h(\tilde v_i) - p(b_i, \tilde v_i) h'(\tilde v_i)}{h^2(\tilde v_i)} \opd \tilde v_i.
\end{align*}
Based on the monotonicity of equilibrium (\lref{lem:exists}), the assumptions on the payment function $p$ (cf. \sref{sec:model}) imply that $p(b,v)>0$ for any $b>0=b(\uline v)$. Furthermore, as $h'(v)>0, p'_v(b,v)\le 0$ and $h(v)\ge 0$, we conclude that $u_i\ge 0$, which proves IR, and that the equality holds iff $v_i=\uline v$ (which subsumes the case of $v_i=0$ since $\uline v\ge 0$). However, an agent of $v_i=\uline v$ will choose not to participate ($b_i=0$) as explained in the proof of \lref{lem:strategy}. Therefore, any agent who exerts nonzero effort reaps a strictly positive payoff.
\Endproof

\subsubsection{Incentive Compatibility (IC)}
We say a (direct revelation) mechanism satisfies IC if agents will report their types truthfully. In our mechanism, prize allocation is based on agents' {\em observable} efforts and the common prior about all the agents' types, instead of on reported (if any), unobservable types. Therefore, the issue of truthful type-reporting is technically irrelevant to our mechanism.
\metacom{check [duan12infocom,online14tpds]!}

\section{Case Study}\label{sec:simu}
\metacom{tabulate key formulas to demonstrate useful results!}

In order to derive an intuitive understanding and draw further insights, we provide a numerical case study that involves six mechanisms: OPT, FIX, SYM-1, SYM-2, OPT-n, and FIX-n.

{\bf OPT} is instantiated with two agents of types $v_1,v_2\in[0,1]$ which are independently drawn from $F_1(v)=v$ (uniform distribution) and $F_2(v)=\frac{v +1}{2}$, respectively. Hence, $f_{2}(v)=\inv{2} \delta(v) + \inv{2}$ where $\delta(\cdot)$ is the Dirac delta function.\footnote{Neither our model nor analysis assumes continuity of the p.d.f. at the boundary of the support, and hence our results still apply. We also chose a power-law distribution $F_2(v)=v^\alpha, \alpha>0$, in the comparison and obtained similar results; however, the actual expressions are too long (due to the inverse effort function $v_2(b_2)$) to suit a neat presentation and hence omitted.} That is, agent 2 is of type zero with probability 0.5, and draws his type uniformly from (0,1] with probability of the other half; so, agent 1 is statistically {\em stronger} than agent 2.
The value function $V(v, Z)=vZ$ and the payment function is $p(b,v)=b^2$. Hence, $h(v)=v$ and $\hat p(b,v)=b^2/v$.

We compare OPT with all counterpart mechanisms which are three canonical auctions:
\begin{itemize}
\item {\bf FIX}: Fixed-prize asymmetric all-pay auctions.
\item {\bf SYM}: Fixed-prize symmetric all-pay auctions, including
\begin{itemize}
\item {\bf SYM-1}: both types follow $F_1(v)$;
\item {\bf SYM-2}: both types follow $F_2(v)$.
\end{itemize}
\end{itemize}

In order to investigate how SA neutralizes the law of DMR and enhances scalability, which requires a larger-scale simulation, we compare {\bf OPT-n} and {\bf FIX-n}, which are OPT and FIX both with $n$ symmetric agents (choosing agent 1 for illustration).

\subsection{Theoretical underpinnings}

To carry out the comparison, we need the following analytical results for FIX and SYM. The proofs are available in \cite{mass14app}.
\ifdefined\JNL \newcommand{\Z}{\blue Z} \else \newcommand{\Z}{} \fi
\ifdefined\JNL \newcommand{\sqrtZ}{\blue{\sqrt Z}} \else \newcommand{\sqrtZ}{} \fi
\begin{prop}[Equilibrium strategy in FIX]\label{thm:strategy-asym2FP}
In a two-player asymmetric all-pay contest with incomplete information, if the common prior is $F_1(v),F_2(v),v\in[\uline v, \bar v]$, \ifdefined\JNL \blue{the value function is $V(v,Z)=vZ$ for any prize $Z$,} \fi and the payment function $p(b)$ satisfies $p(0)=0$ and $p'(b)>0$, there exists a unique Bayesian Nash equilibrium $\vect b=(b_1,b_2)$ which is given by
\begin{align}
b_1(v_1) &= p^{-1} \Big(\Z \int_{k^{-1}(\uline v)}^{v_1} k(v) F'_1(v) \opd v \Big), \label{eq:strategy1-asym2FP}\\
b_2(v_2) &= b_1( k^{-1}(v_2) ),
\end{align}
where $b_1(v)=0$ iff $v_1=k^{-1}(\uline v)$, and $k(v)$ is determined by
\[ k'(v) = \frac{k(v) F'_1(v)}{v F'_2( k(v) )} \]
with boundary condition $k(\bar v)=\bar v$.
\end{prop}\metafoot{Proof is in the spirit of \cite{amann96asym}. Our model can be seen as a generalization of a standard (fixed-prize) asymmetric all-pay auctions. Therefore \pref{thm:strategy-asym2FP} should be able to be derived as a corollary of \lref{lem:strategy}. Try? In addition, the reason why not just use \lref{lem:strategy} directly is that it seems to be still too ``implicit'' for a fairly convenient numerical calculation.}

\begin{prop}[Equilibrium strategy in SYM]\label{thm:strategy-symFP}
In a $n$-player symmetric all-pay auction with incomplete information, if the common prior is $F(v),v\in[\uline v, \bar v]$, \ifdefined\JNL \blue{the value function is $V(v,Z)=vZ$ for any prize $Z$,} \fi and the payment function $p(b)$ satisfies $p(0)=0$ and $p'(b)>0$, there exists a unique Bayesian Nash (symmetric) equilibrium which is given by
\begin{align}
b(v) = p^{-1} \Big( v\Z F^{n-1}(v) - \Z \int_{\uline v}^v F^{n-1}(t) \opd t \Big).
\end{align}
\end{prop}

\subsection{Agent strategy, Contest prize, and Principal's profit}
\subsubsection{Agent strategy}~\\
{\bf OPT:}
Using \thmref{thm:opt-prize}, we apply \eqref{eq:opt-effort} with $\hat p(b_i,v_i)=b_i^2/v_i$ and $F_1=v_1$ to obtain
\[ \frac{2 b_1}{v_1} = \inv \lambda - \frac{2 b_1}{v_1^2} (1-v_1), \]
which gives the optimal equilibrium strategy for agent 1:
\begin{align}\label{eq:strategy-agt1}
b_1(v_1) = \frac{v_1^2}{2 \lambda}, \qquad v_1(b_1) = \sqrt{2 \lambda b_1}.
\end{align}

Similarly, applying \eqref{eq:opt-effort} with $F_2=\frac{v_2+1}{2}$ yields for agent 2:
\begin{align}\label{eq:strategy-agt2}
b_{2}(v_{2}) = \frac{v_{2}^2}{2 \lambda}, \qquad v_{2}(b_{2}) = \sqrt{2 \lambda b_{2}}
\end{align}
which is the same (i.e., symmetric) as agent 1. 
This is because the two type distributions happen to have identical {\em hazard rate}\cite{myerson81}, $\frac{f(v)}{1-F(v)}$, which is used (inversely) in \eqref{eq:opt-effort}. Clearly, this should not be generalized to all c.d.f.'s; indeed, we shall see later that the optimal prizes for the two agents \eqref{eq:prize-agt1}\eqref{eq:prize-agt2} as well as their individual contributions to the principal's profit \eqref{eq:prof1-opt}\eqref{eq:prof2-opt} are different.

{\bf FIX:}
Instantiating \pref{thm:strategy-asym2FP} with $F_1(v)=v$ and $F_2(v)=\frac{v+1}{2}$ yields
\begin{align}
k'(v) = \frac{2 k(v)}{v} &\Rightarrow k(v) = v^2, \qquad k^{-1}(v)=\sqrt{v}. \nn\\
\text{Therefore}\quad b^{fix}_1(v_1) &= \Big(\Z\int_0^{v_1} v^2 \opd v\Big)^{\inv 2} 
= \frac{v_1^{3/2}}{\sqrt 3}\sqrtZ, \label{eq:strategy-FP-agt1} \\
b^{fix}_{2}(v_2) &= \frac{v_2^{3/4}}{\sqrt 3}\sqrtZ. \label{eq:strategy-FP-agt2}
\end{align}

{\bf SYM:}
For SYM-1, applying \pref{thm:strategy-symFP} with $F(v)=v$ gives
\begin{align}\label{eq:strategy-sym-agt1}
b_1^{sym}(v) = \frac{v}{\sqrt 2}\sqrtZ.
\end{align}
For SYM-2, applying \pref{thm:strategy-symFP} with $F(v)=\frac{v+1}{2}$ gives
\begin{align}\label{eq:strategy-sym-agt2}
b_{2}^{sym}(v) = \frac{v}{2}\sqrtZ.
\end{align}

\subsubsection{Optimal prize tuple in {\bf OPT}}~\\
\thmref{thm:opt-prize} gives the optimal prize for agent~1 via \eqref{eq:opt-prize}:
\begin{align}\label{eq:prize-agt1}
\hspace{-3mm}Z_1(b_1) = \frac{ \frac{v_1^3}{4 \lambda^2} + 
\int_0^{v_1} \frac{\tilde v_1^2}{4 \lambda^2} \opd \tilde v_1 }
                            {\frac{v_1+1}{2}} \Big|_{v_1=\sqrt{2 \lambda b_1}}
= \frac{ 2 (\sqrt{2 \lambda b_1})^3 } {3 \lambda^2 (\sqrt{2 \lambda b_1} + 1) }.
\end{align}
The optimal prize for agent 2 is similarly obtained to be
\begin{align}\label{eq:prize-agt2}
Z_{2}(b_{2}) = \frac{2}{3\lambda} b_{2}.
\end{align}

\subsubsection{Principal's profit}~\\
{\bf OPT:}
The principal's maximized profit can again be calculated using \thmref{thm:opt-prize}. Particularly in \eqref{eq:profit-final}, we calculate each agent's contribution corresponding to each summation term of $\pi=\pi_1 + \pi_2$, as follows:
\begin{align}
\pi_1 &= \int_0^1 
\big[\frac{v^2}{2 \lambda} - \frac{v^3}{4 \lambda} - \frac{v^2}{4 \lambda} (1-v) \big] \opd v
= \inv{12 \lambda}, \label{eq:prof1-opt}\\
\pi_{2} &= \int_{0^+}^1 
\big[\frac{v^2}{2 \lambda} - \frac{v^3}{4 \lambda} - \frac{v^2}{4 \lambda} (1-v) \big] \frac{\opd v}{2} = \inv{24 \lambda}.\label{eq:prof2-opt} \\
\therefore \pi &= \pi_1 + \pi_{2} = \inv{8 \lambda}.\label{eq:profit-var}
\end{align}
In the above when calculating $\pi_{2}$, although there is a probability atom of 0.5 at $v=0$, the effort and payment are both zero, and hence it does not contribute to the profit and we can take the integral from $0^+$ onward. 

{\bf FIX:}
The profit in this case is $\pi^{fix} = \pi^{fix}_1 + \pi^{fix}_{2} - \lambda\Z $ where
\begin{align}
\pi^{fix}_1 &= \int_0^1 b^{fix}_1(v_1) \opd F_1(v_1) = \frac{2\sqrtZ}{5 \sqrt 3},\nn\\
\pi^{fix}_{2} &= \int_{0^+}^1 b^{fix}_{2}(v_2) \opd F_2(v_2) = \frac{2\sqrtZ}{7 \sqrt 3}.\nn\\
\therefore \pi^{fix} &= \frac{24\sqrtZ}{35 \sqrt 3} - \lambda\Z. \label{eq:profit-fix}
\end{align}
Like in OPT, $b^{fix}_{2}(0)=0$ nullifies the atom at $f_2(0)$.

\ifdefined\JNL 
\blue{New:} The profit maximizing problem $\max_{Z} \pi^{fix}$ can be easily solved using FOC: $\sqrt{Z^*}=\frac{24}{35 \sqrt 3} / (2\lambda)$ and 
\[ \pi^{fix^*}=(\frac{24}{35 \sqrt 3})^2 \Big/ (4\lambda) = \frac{48}{1225\lambda}.\]
\fi

\ifdefined\JNL \newcommand{\sqrtZone}{\blue{\sqrt Z}} \else \newcommand{\sqrtZone}{1} \fi
{\bf SYM:}
The profits of SYM-1 and SYM-2 are, respectively,
\begin{align}
\pi_1^{sym} &= 2 \int_0^1 b^{sym}_1(v_1) \opd F_1(v_1) - \lambda\Z = \frac{\sqrtZone}{\sqrt 2} - \lambda\Z, \label{eq:profit-sym1}\\
\pi_2^{sym} &= 2 \int_0^1 b^{sym}_2(v_2) \opd F_2(v_2) - \lambda = \frac{\sqrtZone}{4} - \lambda\Z.\label{eq:profit-sym2}
\end{align}

\ifdefined\JNL 
\blue{New:} Similar to FIX, the profit maximizing problem can be solved using FOC:
\begin{align*}
\sqrt{Z_1^*}&=\inv{2\sqrt 2 \lambda} &\text{ and }&& \pi_1^{sym^*}&=\inv{8\lambda},\\
\sqrt{Z_2^*}&=\inv{8\lambda} &\text{ and }&& \pi_2^{sym^*}&=\inv{64\lambda}.
\end{align*}
\fi

\subsection{Results}

\begin{figure}[t]
\centering
\fbox{\includegraphics[trim=4cm 5.7cm 4.9cm 6cm,clip,width=.75\linewidth]{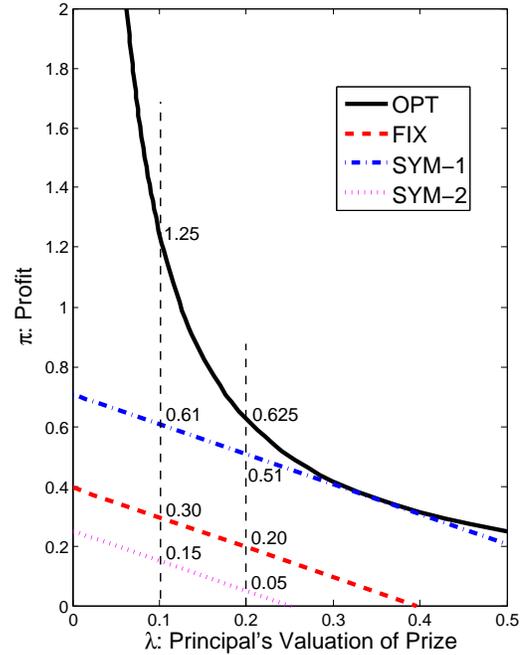}}
\caption{Profit comparison of different mechanisms.}
\label{fig:profit}
\end{figure}

In line with the organizer's ultimate objective, we first compare the profit of the above four mechanisms in \fref{fig:profit}, based on formulae \eqref{eq:profit-var}--\eqref{eq:profit-sym2}.
The plot clearly shows that OPT garners the highest profit compared to all the other mechanisms over all possible $\lambda$. In particular, even though SYM-1 is privileged to benefit from two strong agents, it is still outperformed by OPT, apart from being tangent to OPT at only one point ($\lambda=\sqrt{2}/4$). Specifically, eight profit values are also marked in \fref{fig:profit} at $\lambda=0.1$ and 0.3, where we see that OPT significantly outperforms the other three mechanisms, by about 105\%, 315\% and 730\%, respectively (at $\lambda=0.1$). If $\lambda$ is sufficiently high, FIX, SYM-1, and SYM-2 even run into deficit (negative profit), at $\lambda>0.396$, $\lambda>0.707$, and $\lambda>0.25$, respectively. On the other hand, as $\lambda$ becomes smaller (i.e., the principal values the prize less), OPT reaps {\em exponential} profit growth whereas the other mechanisms only have linear profit increase.

\begin{figure*}[tb]
\centering
\subfloat[$\lambda=0.1$.]{\label{fig:strategy-01}
\fbox{\includegraphics[trim=3.9cm 8.5cm 4.7cm 9cm,clip,width=0.32\linewidth]{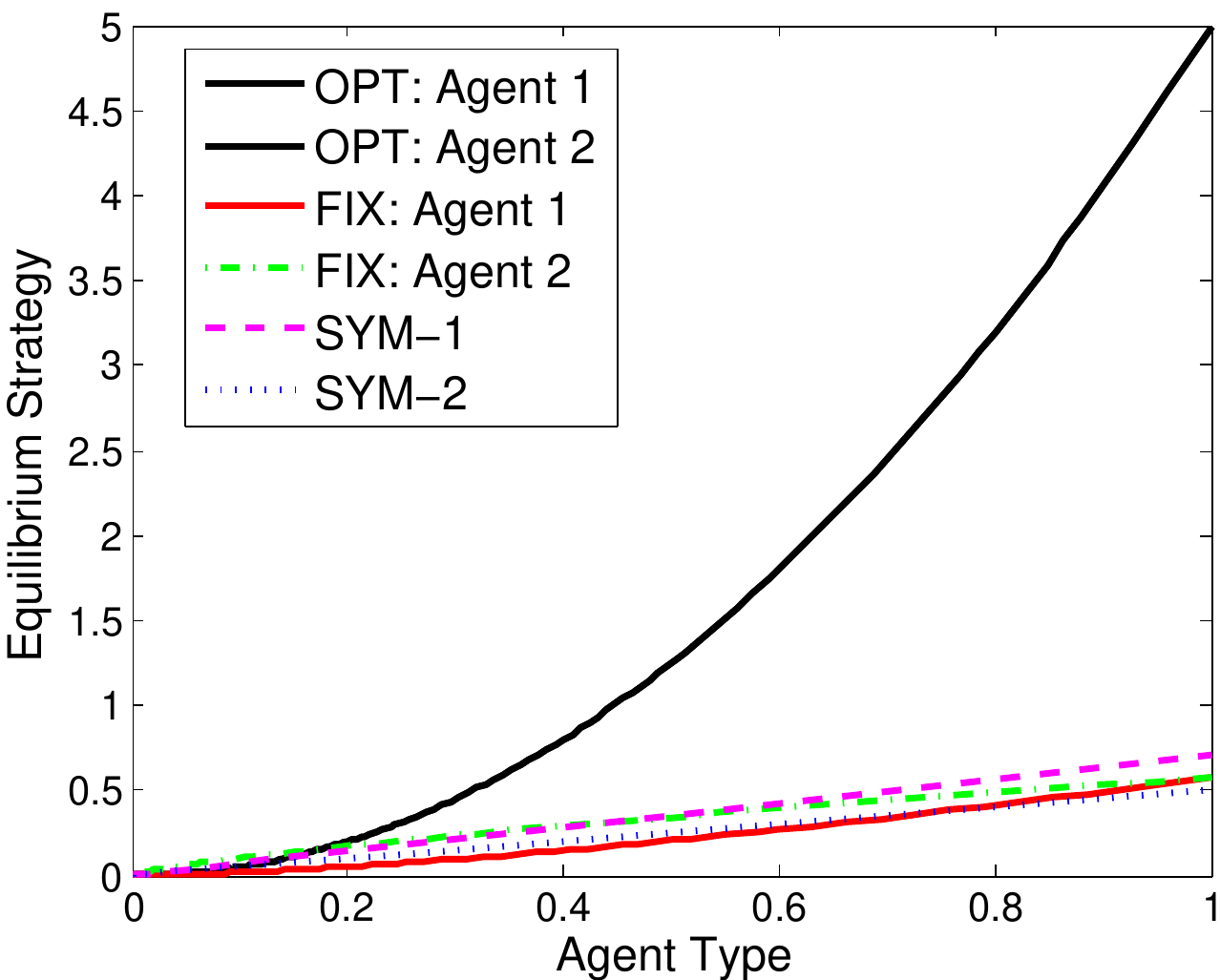}}}\hfil
\subfloat[$\lambda=0.3$.]{\label{fig:strategy-03}
\fbox{\includegraphics[trim=3.9cm 8.5cm 4.7cm 9cm,clip,width=0.32\linewidth]{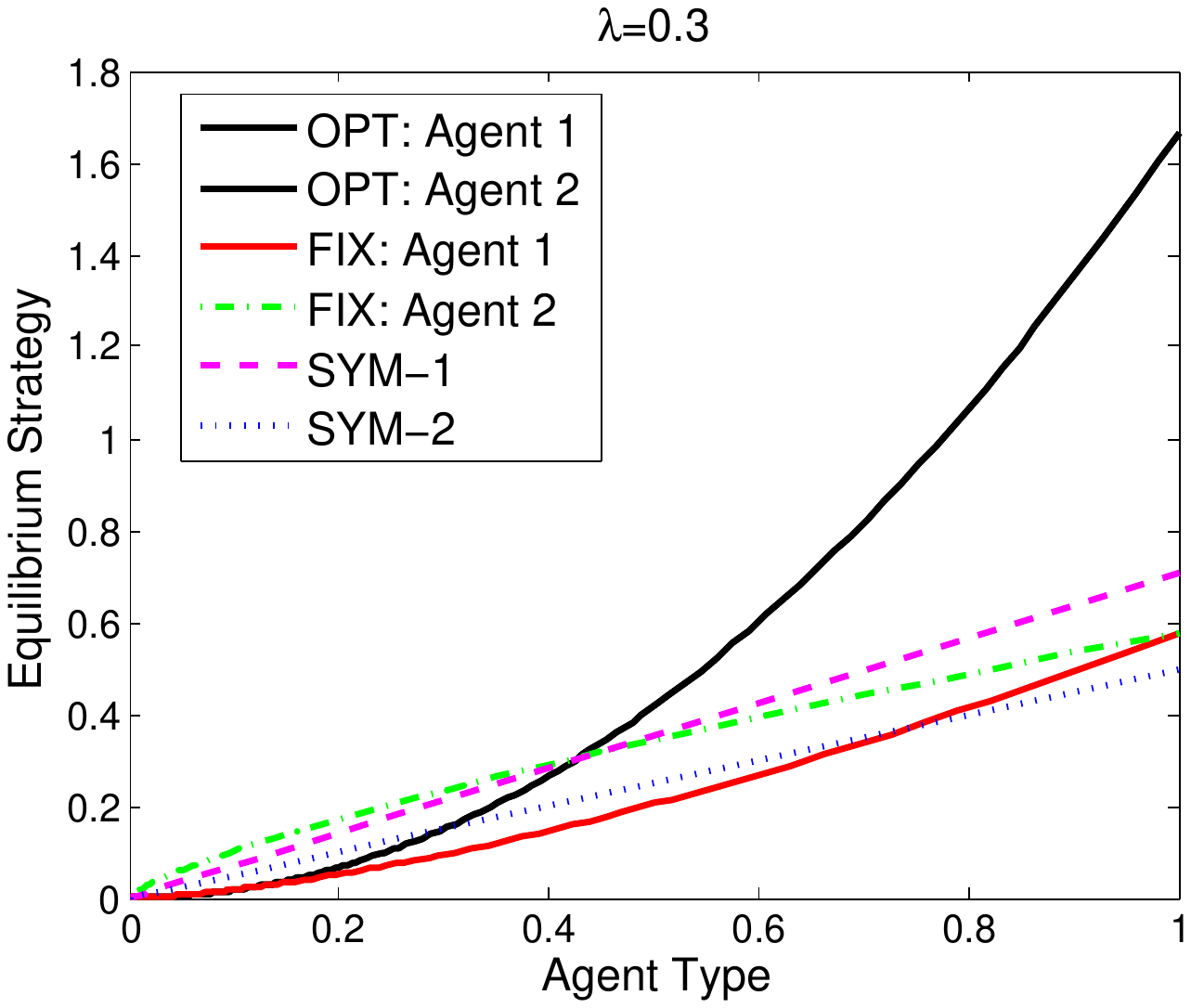}}}\hfil
\subfloat[$\lambda=0.5$.]{\label{fig:strategy-05}
\fbox{\includegraphics[trim=3.9cm 8.5cm 4.7cm 9cm,clip,width=0.32\linewidth]{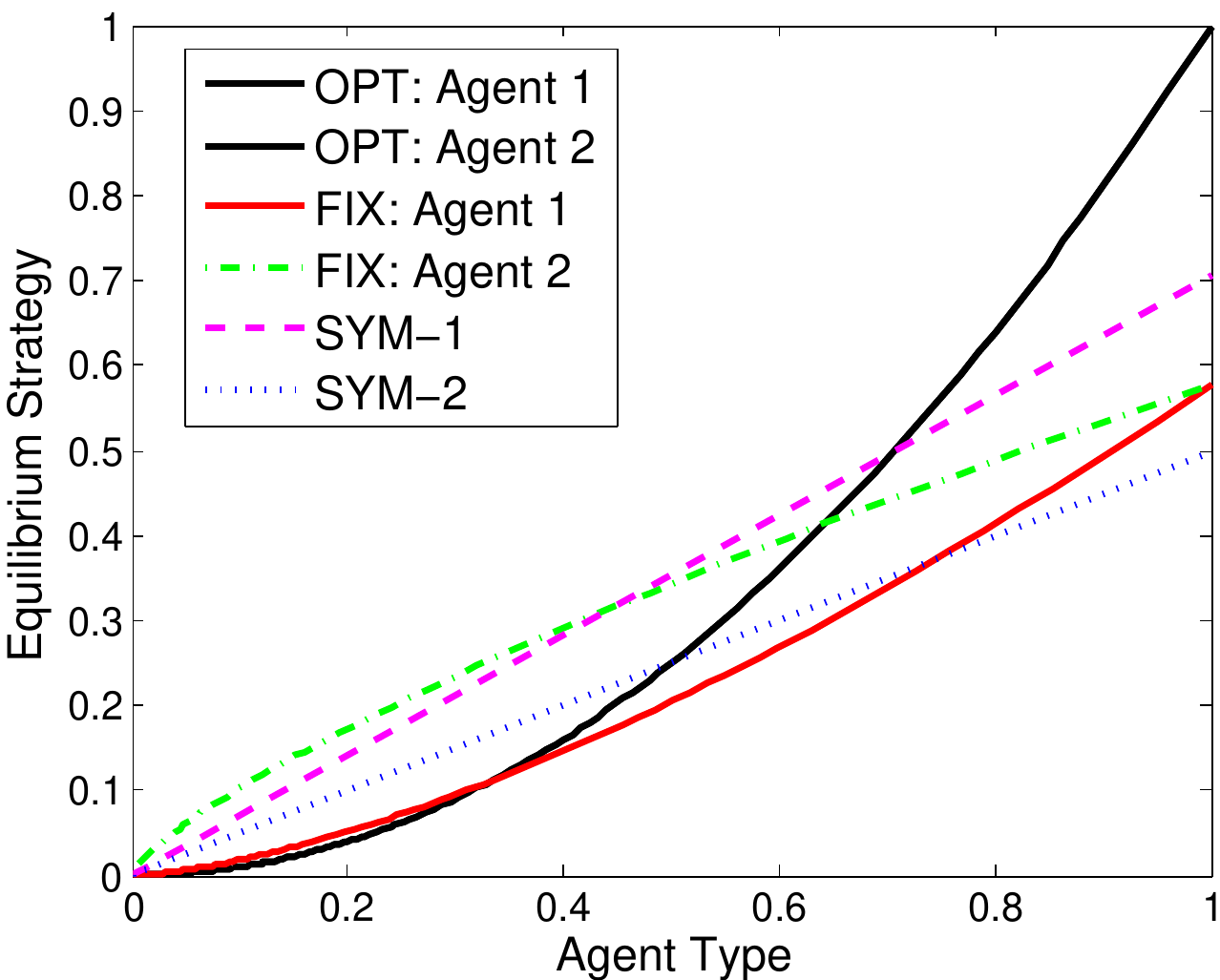}}}
\caption{Equilibrium strategy (agent effort). The same line-spec is used for both agents in OPT as they adopt the same strategy.}
\label{fig:strategy}
\end{figure*}

\begin{figure*}[tb]
\centering
\subfloat[$\lambda=0.1$; the maximum effort is 5.]{\label{fig:prize-01}
\fbox{\includegraphics[trim=3.9cm 8.5cm 4.7cm 9cm,clip,width=0.32\linewidth]{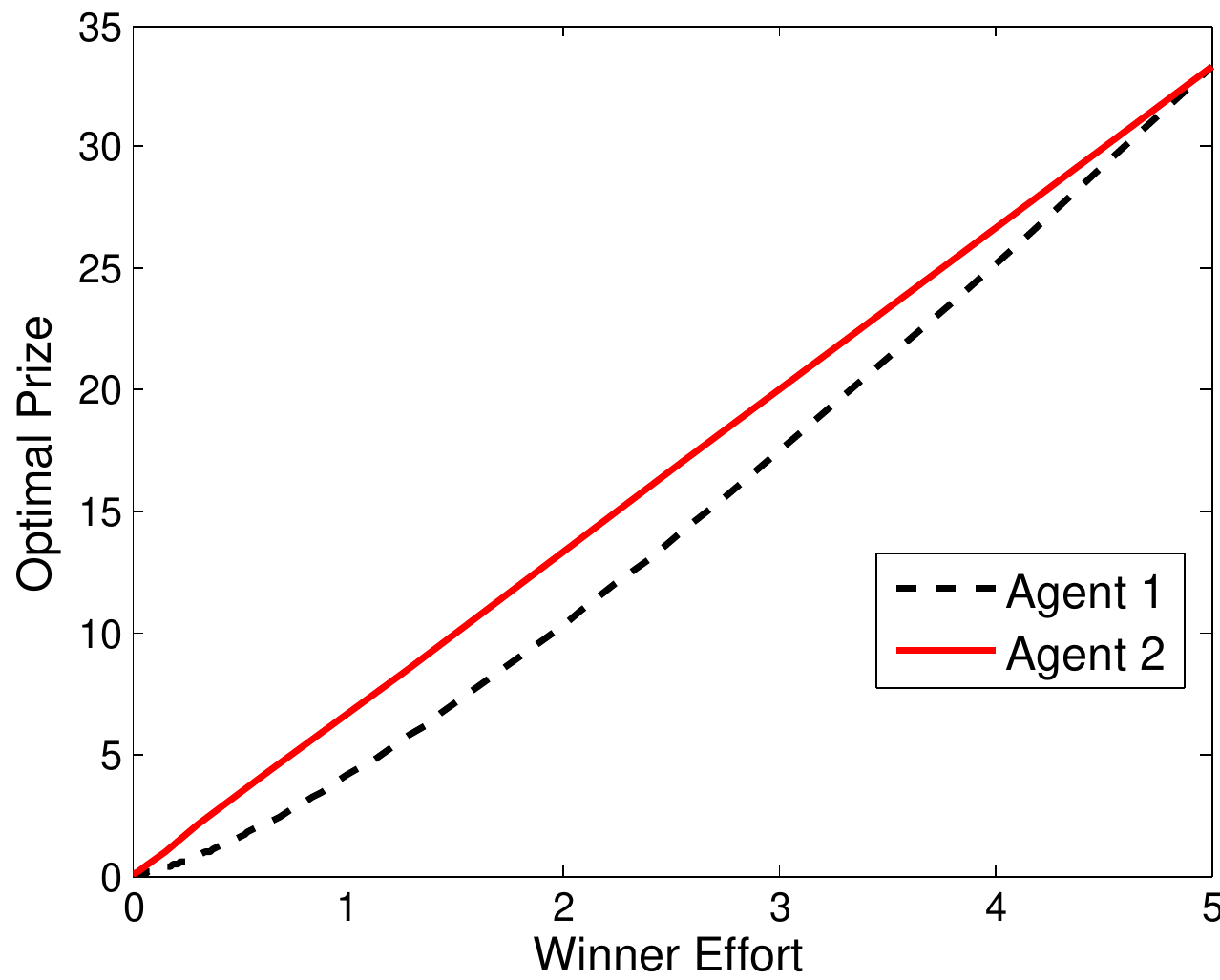}}}\hfil
\subfloat[$\lambda=0.3$; the maximum effort is $\frac{5}{3}$.]{\label{fig:prize-03}
\fbox{\includegraphics[trim=3.9cm 8.5cm 4.7cm 9cm,clip,width=0.32\linewidth]{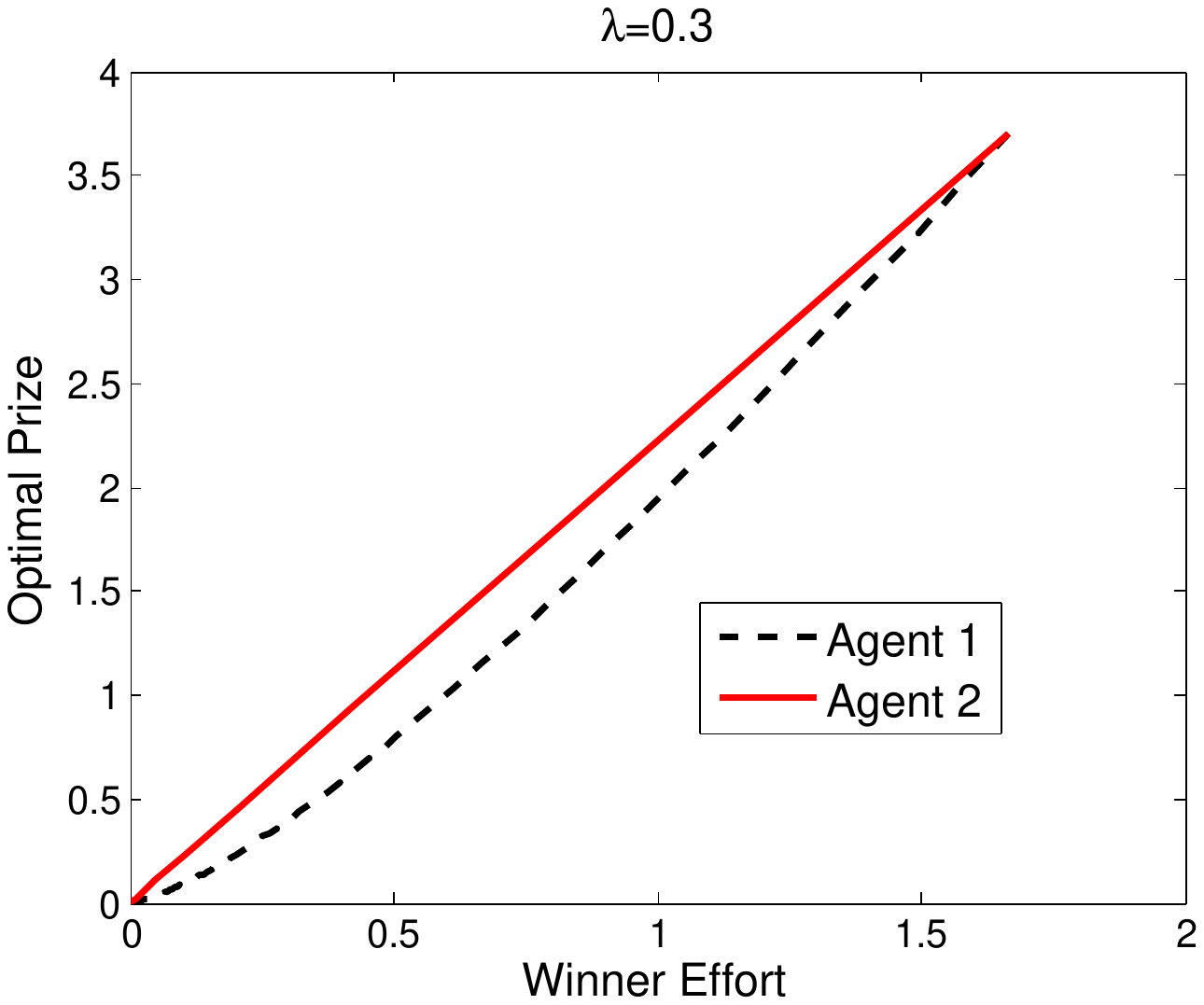}}}\hfil
\subfloat[$\lambda=0.5$; the maximum effort is 1.]{\label{fig:prize-05}
\fbox{\includegraphics[trim=3.9cm 8.5cm 4.7cm 9cm,clip,width=0.32\linewidth]{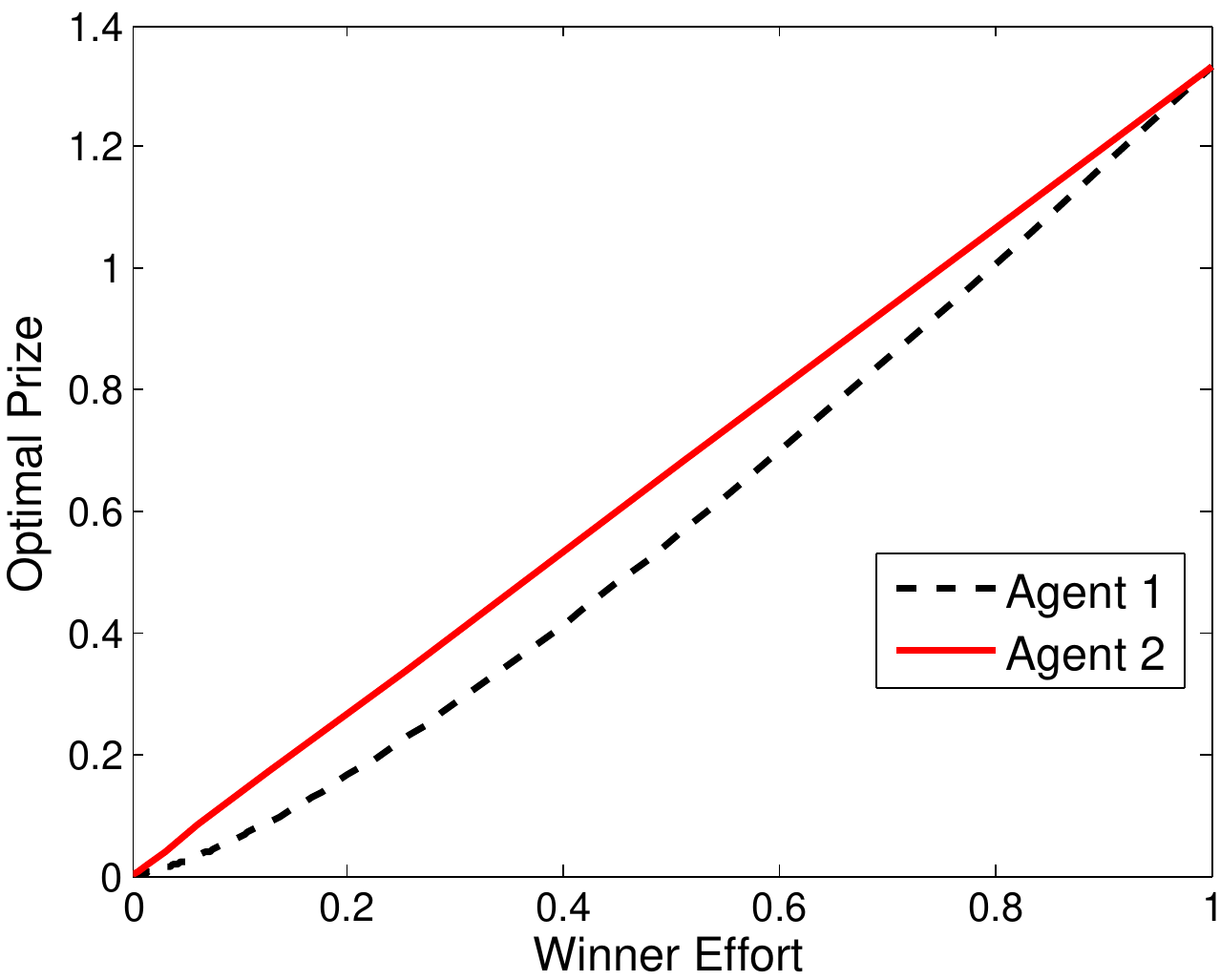}}}
\caption{Optimal prizes as functions of winner efforts in OPT. The range of X axis is determined by the maximum effort.}
\label{fig:prize}
\end{figure*}

\vspace{1mm}
\subsubsection{Profit ranking and rationale}
According to \fref{fig:profit}, the profit of the four mechanisms can be ranked as SYM-2 $\prec$ FIX $\prec$ SYM-1 $\prec$ OPT, where $\prec$ denotes ``is inferior to''. To understand the rationale behind this ranking result, we examine the agent strategies, by plotting formulae \eqref{eq:strategy-agt1}--\eqref{eq:strategy-sym-agt2} in \fref{fig:strategy}.

In general, the ranking SYM-2 $\prec$ FIX $\prec$ SYM-1 can be understood by the composition of the three contests: SYM-1 and SYM-2 are composed of two strong and two weak agents, respectively, and FIX is a mixture. However, it is worth noting that FIX is much lower than the average of SYM-1 and SYM-2; in fact, it is even lower than half of SYM-1 alone. This is due to the {\em effort reservation} effect existing in asymmetric auctions, as mentioned in \sref{sec:property}, where a stronger agent shades his bid when facing a weaker agent. Indeed, \fref{fig:strategy} shows that agent 1 in FIX bids significantly lower than in SYM-1, although agent 2 exerts higher effort than in SYM-2.\footnote{The reason why agent 2 works harder in FIX than in SYM-2 is because he can deduce that the stronger agent will reserve effort and hence he (agent 2) sees a better chance to win by striving above his (usual) effort level as in the symmetric case (SYM-2).}  The reason why the effort reduction of the stronger agent outweighs the effort increase of the weaker agent is that, mathematically, the p.d.f. of the stronger type concentrates on the higher region of the common support $[\uline v, \bar v]$ and thus has a larger impact on the revenue (calculated by an integral). Intuitively, this tells that ``stronger agents matter more'', and offers us the following insight: it is {\em more productive} for a mechanism to focus on incentivizing {\em stronger} agents who constitute the main contributors to the revenue. This hints toward a {\em discriminatory} contest design, and indeed, this discriminatory design principle is used by both this work (the agent-specific prize functions) and some prior work such as \cite{discrim98}.

To understand why SYM-1 $\prec$ OPT, first we see in \fref{fig:strategy} that OPT incentivizes agents to exert significantly higher effort than all the other mechanisms, particularly at higher types, which concurs the productivity of incentivizing stronger agents mentioned above.

To draw deeper insights into how this is achieved, we examine the optimal prize tuple of OPT, by plotting formulae \eqref{eq:prize-agt1}\eqref{eq:prize-agt2} in \fref{fig:prize}. We see that OPT gives slightly higher reward to agent 2 if he exerts the same amount of effort as agent 1. The rationale is to motivate the weaker agent insofar as he becomes a {\em competitive rival} to the stronger agent, thereby ``threatening'' the stronger agent {\em not} to reserve effort. In principle, the prize tuple endogenizes agent asymmetry and enables the contest to recuperate from the fierceness of competition existing in symmetric contests.
Moreover, the increasing monotonicity of the prize functions also motivates agents to work harder as the reward grows with their effort. These incentives render OPT superior to SYM-1 and the other mechanisms.

Lastly, as a side note, \fref{fig:strategy} also indicates that the agent strategy in all the mechanisms is monotone increasing in agent type, which conforms to \lref{lem:exists}. 

\begin{figure*}[tb]
\centering
\subfloat[$\lambda=0.1$.]{\label{fig:nsym-01}
\fbox{\includegraphics[trim=4.3cm 8.2cm 4.7cm 8.8cm,clip,width=0.32\linewidth,height=4.7cm]{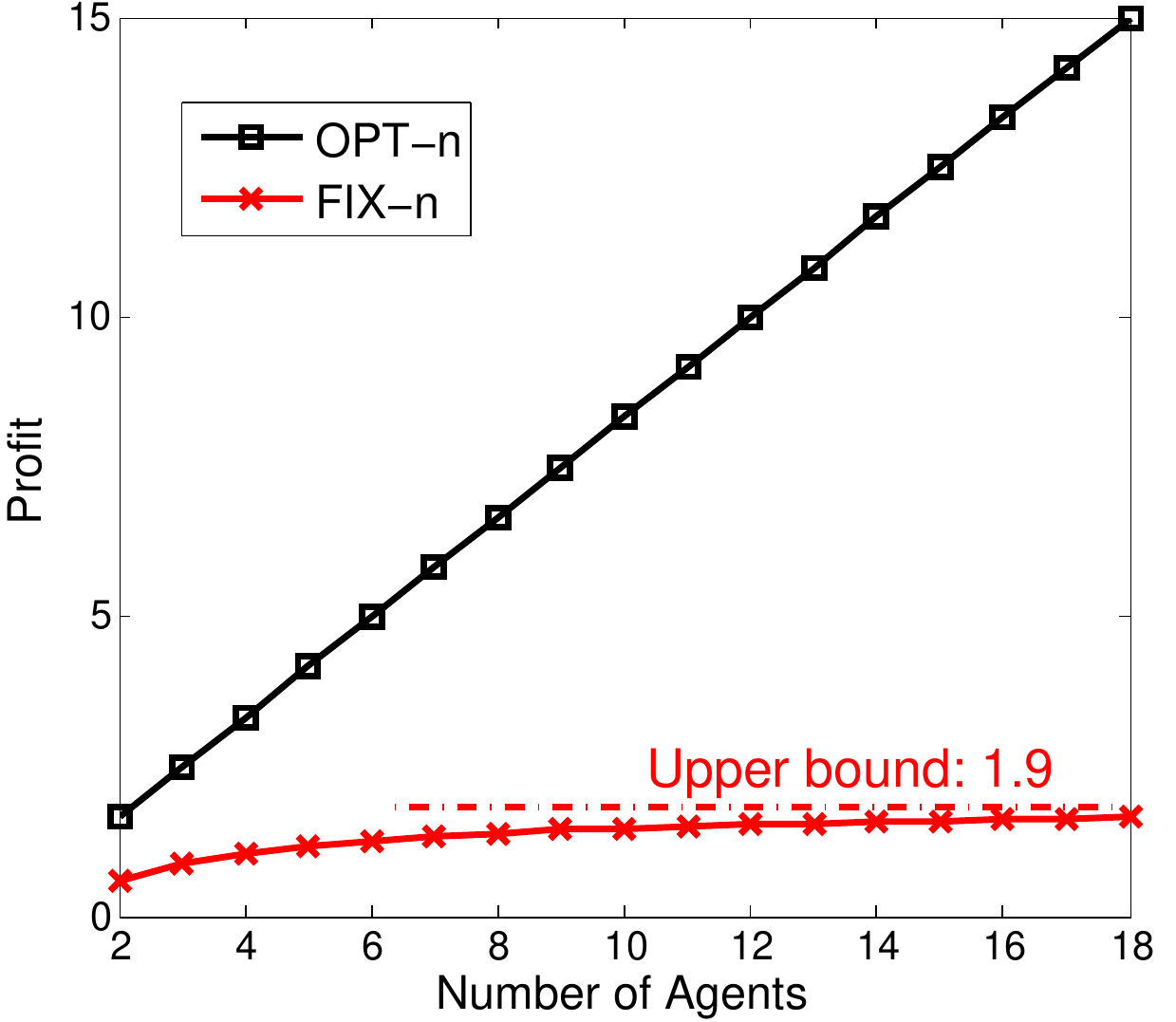}}}\hfil
\subfloat[$\lambda=0.3$.]{\label{fig:nsym-03}
\fbox{\includegraphics[trim=3.9cm 8.5cm 4.5cm 9cm,clip,width=0.32\linewidth]{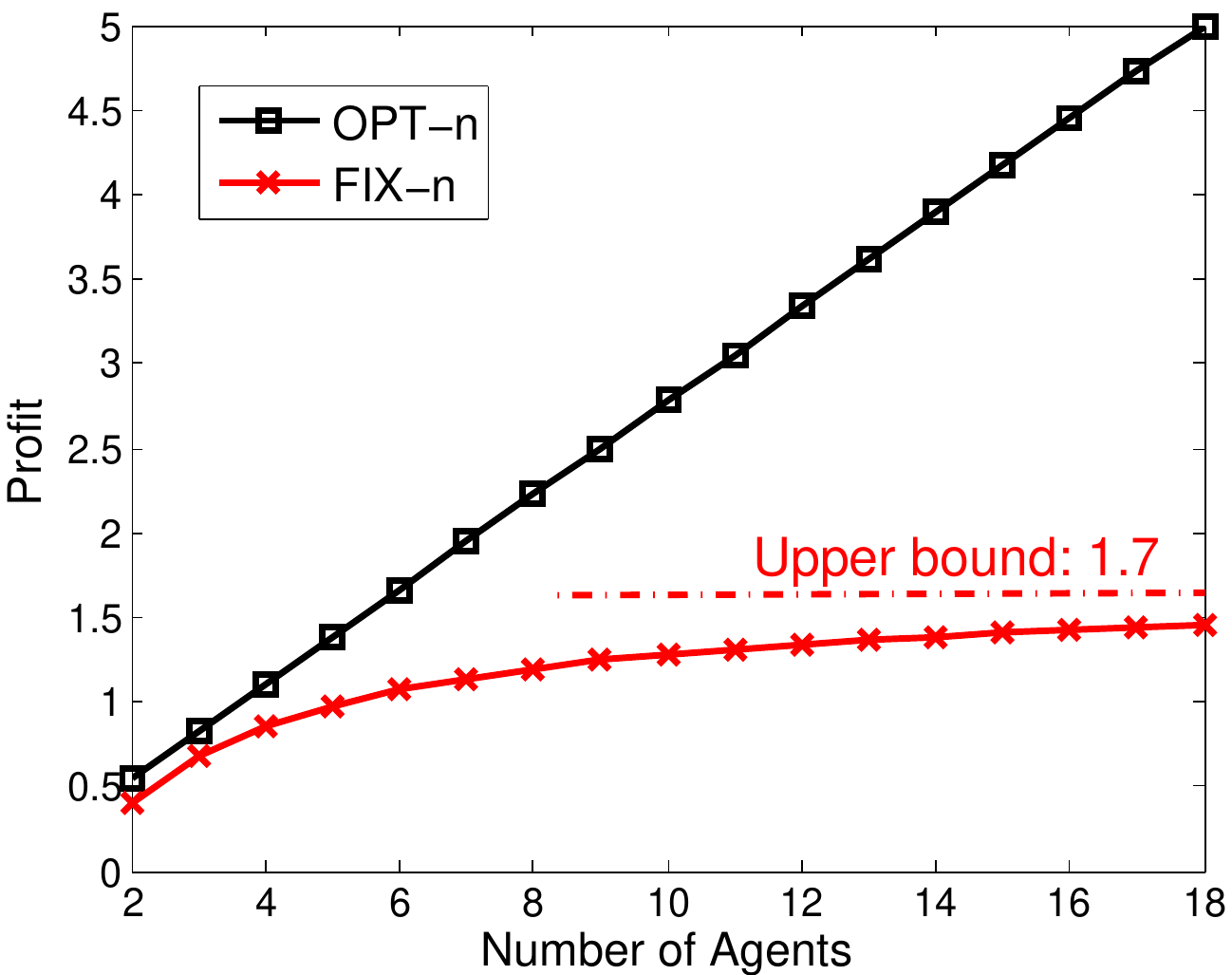}}}\hfil
\subfloat[$\lambda=0.5$.]{\label{fig:nsym-05}
\fbox{\includegraphics[trim=3.9cm 8.5cm 4.5cm 9cm,clip,width=0.32\linewidth]{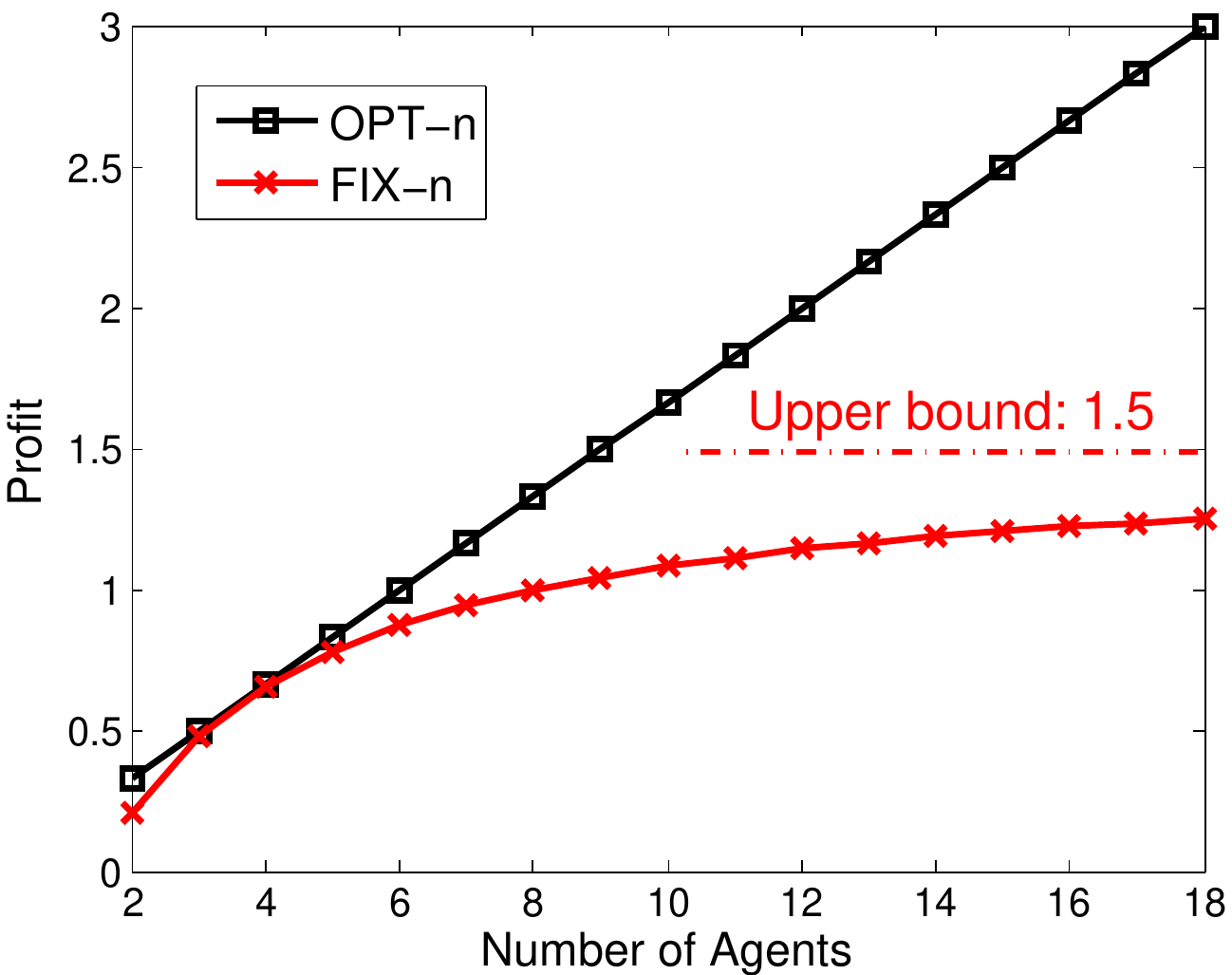}}}
\caption{Strategy autonomy neutralizes the law of diminishing marginal returns.}
\label{fig:nsym}
\end{figure*}

\vspace{2mm}
\subsubsection{Neutralizing DMR}\label{sec:DMRsimu}

In this subsection, we investigate how SA neutralizes the law of DMR, by comparing OPT-$n$ and FIX-$n$. In OPT-$n$, since the $n$ agents are now homogeneous (same as agent 1, for illustration), the prize tuple collapses into a single prize function. Using \thmref{thm:opt-prize}, this single, optimal prize function can be calculated as\metafoot{Short cut is not to use theorem 1 but use \eqref{eq:prize-agt1}.}
\[ Z^{opt\text-n}(b) = \frac{(2 \lambda b)^{2-\frac{n}{2}} }{3 \lambda^2}.\]
In addition, the equilibrium agent strategy becomes
\begin{flalign}
&b = \frac{v^2}{2 \lambda},&\label{eq:optn-strategy}\\
\noindent\text{and the resultant profit is}\qquad &\pi^{opt\text-n} = \frac{n}{12 \lambda}.\label{eq:optn-profit}
\end{flalign}

For FIX-$n$, the equilibrium agent strategy is calculated using \pref{thm:strategy-symFP}, as
\begin{align}\label{eq:fixn-strategy}
b^{fix\text-n}(v) = \sqrt{\frac{n-1}{n}} v^{\frac{n}{2}},
\end{align}
and the resultant profit is
\begin{align}\label{eq:profit-fixn}
\hspace{-3.5mm}\pi^{fix\text-n} = n \int_0^1 b^{fix\text-n}(v) \opd F_1(v) - \lambda
= \frac{2 \sqrt{n(n-1)}}{n+2} - \lambda 
\end{align}

We plot $\pi^{opt\text-n}$ and $\pi^{fix\text-n}$ in \fref{fig:nsym} with respect to $n$ for different $\lambda$ values. As we can see, as a standard auction, FIX-$n$ is indeed governed by the law of DMR, and exhibits concave profit growth as $n$ increases. Eventually, it saturates at the upper bound of  $\lim_{n\to\infty} \pi^{fix\text-n} = 2 - \lambda$, which is indicated in \fref{fig:nsym} as well.

On the contrary, OPT-$n$ is not confined by the law of DMR and its profit grows {\em linearly} as $n$ increases. To understand why, we note that in the symmetric case, SA is reinterpreted as that the agent strategy is independent of the number of agents. This is also evidenced by \eqref{eq:optn-strategy}, whereas the strategy in FIX-$n$ \eqref{eq:fixn-strategy} does depend on $n$. Therefore the revenue---the sum of all the agents' bids---is a linear function of $n$ (in detail, revenue is $n \int_0^1 \frac{v^2}{2 \lambda} \opd F(v) = \frac{n}{6\lambda}$). The cost is also linear in $n$: $\lambda \mathbbm E[Z] = \lambda \int_0^1 \frac{v^{4-n}}{3 \lambda^2} \opd v^n = \frac{n}{12 \lambda}$. Therefore, the profit is a linear function of $n$, verified by $\frac{n}{6\lambda} - \frac{n}{12 \lambda} = \frac{n}{12 \lambda}$ which coincides with \eqref{eq:optn-profit}.

Moreover, when $n$ is not too small, as in most real scenarios, \fref{fig:nsym} shows that OPT-$n$ garners much larger profit which is even superior to the {\em upper bound} of FIX-$n$ by far. These observations manifest a very healthy {\em scalability} for OPT-based crowdsourcing systems.

\section{Conclusion}\label{sec:conc}

This paper frames the problem of incentive mechanism design for crowdsourcing into an all-pay contest model. To the best of our knowledge, our model represents the first contest or auction model that (a) accommodates multiple heterogeneous users with incomplete information, and (b) is instrumented with a prize function tuple as opposed to the conventional, single, fixed prize. Not only does it more closely characterizes realistic crowdsourcing campaigns, but it also induces the highest possible total effort from self-interested agents, which is of utmost importance to most crowdsourcing campaigns.

The strategy autonomy (SA) property, which is discovered during the course of investigating this new model, captures a counter-intuitive and surprising phenomenon: agents in a heterogeneous environment behave independently of one another as if they were in a homogeneous one.  SA bears practical significances pertaining to system complexity, profitability, and scalability. It could also be an enrichment to the theory of mechanism design.

\bibliography{IEEEabrv}

\end{document}